\documentclass[journal]{IEEEtran}
\usepackage{hyperref}

\usepackage{amsmath,graphicx}
\usepackage{epstopdf}

\usepackage{xcolor}
\usepackage{algorithm} 
\usepackage{algpseudocode}
\usepackage{float,placeins}

\title{Numerically robust tetrahedron-based tomographic forward and backward projectors on parallel architectures}
\author{Ander~Biguri,~Hossein~Towsyfyan,~Richard~Boardman,~Thomas~Blumensath\thanks{A. Biguri, H. Towsyfyan and T. Blumensath are with Institute of Sound and Vibration Research (ISVR), University of Southampton}
\thanks{R. Boardman is with $\mu$-VIS X-Ray Imaging Centre, University of Southampton}
}

\begin{document}
%
\maketitle
\begin{abstract}
X-ray tomographic reconstruction typically uses voxel basis functions to represent volumetric images. Due to the structure in voxel basis representations, efficient ray-tracing methods exist allowing fast, GPU accelerated implementations. Tetrahedral mesh basis functions are a valuable alternative to voxel based image representations as they provide flexible, inhomogeneous partitionings which can be used to provide reconstructions with reduced numbers of elements or with arbitrarily fine object surface representations. We thus present a robust parallelizable ray-tracing method for volumetric tetrahedral domains developed specifically for Computed Tomography image reconstruction. Tomographic image reconstruction requires algorithms that are robust to numerical errors in floating point arithmetic whilst typical data sizes encountered in tomography require the algorithm to be parallelisable in GPUs which leads to additional constraints on algorithm choices. Based on these considerations, this article presents numerical solutions to the design of efficient ray-tracing algorithms for the projection and backprojection operations. Initial reconstruction results using CAD data to define a triangulation of the domain demonstrate the advantages of our method and contrast tetrahedral mesh based reconstructions to voxel based methods.
\end{abstract}
%
%
\section{Introduction}
\label{sec:intro}
Pixels (or voxels in the 3D case) are the most common basis functions used for image representation, their regular grid structure is advantageous for both hardware and software. However, there are alternative bases to represent a continuous scalar field. An alternative are triangular (or tetrahedral) representations. These are heavily used in computer graphics and physical modelling (e.g. in finite element analysis). A tetrahedron-based image representation has desirable properties over regular pixels. They allow for arbitrary spatial  variation in image resolution and can provide arbitrarily precise object boundaries. These properties can be of interest in some applications such as computed tomography (CT), as they can reduce the heavy computational burden of high resolutions scans, improve the conditionality of the reconstruction problem and allow for improved surface models to be derived  from the data. In order to explore the potential of tetrahedral meshes for CT reconstruction, the basic operations of any CT algorithm need to be implemented --- the projection and backprojection operations. These will allow for efficient simulation and reconstruction of CT scans using the proposed image basis. However implementing computationally efficient algorithms for tetrahedral CT comes with significant challenges. 

Firstly, these type of meshes are unstructured, which means that standard computational and numerical approaches typically used to accelerate of CT image reconstruction\cite{chou2011fast}\cite{zinsser2013systematic} no longer apply, as most of them exploit the regular grid structure of voxels to implement efficient numerical algorithms. Secondly robust numerical methods that are not affected by the discrete nature of floating point representations of the data are required, as the image can be represented with units that can significantly vary in size within the same mesh. It is thus important that arithmetic errors will not surpass the required precision to robustly obtain  geometric parameters when computing the required path-integrals. This means that geometric limits of the triangulation need to be set. Finally, due to the data size encountered in many CT problems (especially when using high resolution industrial and micro-CT systems), computationally efficient algorithms are required that can be parallelised over multiple GPUs.

We are thus interested in the development of an algorithm that is as widely applicable as possible, i.e. that puts no or few constrains on the triangulation of the domain (that is, we do not want to restrict our mesh to have tetrahedra with limited aspect ratios, or meshes with triangle density constrains). Furthermore, we want the method to work with any CT reconstruction algorithm, regardless of the mathematical methods used.

Prior research in this field is not extensive. Brankov et al. \cite{brankov2004tomographic} and later Sitek et al.\cite{Sitek2006} proposed using tetrahedra as image basis for Positron Emision Tomography, showing improved image quality over pixels. These results are however restricted to 2D tomography, which numerically is considerably more robust. Yamanaka et al.\cite{Yamanaka2013} proposed a surface reconstruction model for CT using tetrahedral images as basis elements, however their method is only valid for relatively regular tetrahedral meshes of a size not much larger than the detector pixel size. The used reconstruction methods are also limited, as their approach was only valid for the FDK algorithm. Quinto's PhD thesis\cite{Quinto2013} explores triangular and tetrahedral base for  images, and briefly describes a GPU algorithm in a related journal article\cite{Quinto2013TetrahedralVR}. However they use tetrahedra with limited aspect ratios and the method described in their work is thus not applicable to arbitrary triangulations and was found to be prone to failure due to numerical errors when the aspect ratios of the tetrahedra increases. 

We thus propose a numerically robust, parallelisable forward and backward projector for X-ray absorption CT based on ray-tracing. Our method has no constrains on the triangulation and is applicable to arbitrary reconstruction algorithms. We provide a GPU implementation of the method using the CUDA language. The following article will first provide more detail on the numerical problems that may arise in mesh based ray-tracing and  propose robust solutions that allow efficient  GPU implementations. Numerical results  highlight the need for our numerical approach. Results contrasting the difference in tomographic reconstructions achieved with mesh based and voxel based methods highlight potential advantages of the mesh based representations, further motivating our novel approach.

\section{Methods}
\label{sec:methods}
In this section the relevant CT background is introduced and techniques for the calculation of tetrahedron-ray intersection are discussed. A robust numerical method for forward and backpropagation is derived, followed by a discussion of technical aspects of accelerating the method on GPUs. 
\subsection{Computed Tomography}
CT reconstruction is an ill posed inverse problem that can be solved using a wide variety of algorithms. The most common approach is the so-called FDK\cite{FDK} algorithm that consist of two steps: high pass filtering of the measured data and backprojection of the result. Alternatively, the problem can be posed as a linear system of equations of the form 
\begin{equation}
\textbf{A}\textbf{x}=\textbf{b}+\tilde{\mathbf{e}},
\end{equation}
 where $\textbf{A}$ is a matrix where each entry represents the length of the line-element intersection between a voxel (or tetrahedron) and one of the X ray beams, $\textbf{x}$ is the lexicographically ordered attenuation values associated with each voxel (or tetrahedron), $\textbf{b}$ is the lexicographically ordered measured data and $\tilde{\mathbf{e}}$ accounts for errors   in the  data and the linearization of the problem. This equation can be solves using a wide variety of algorithms that exist in the literature including SART\cite{SART}, CGLS\cite{CGLS}, etc.  All algorithms require the computation of $\textbf{A}\textbf{x}$ (the projection operation) and $\textbf{A}^T\textbf{b}$ (the backprojection operations), with the exception of FDK which only  requires the latter. The projection operator simulates each path of an X-ray beam that is measured by a pixel in the detector and integrate the piecewise attenuation coefficients of the image $x_i$ using the intersection length $a_{ij}$ as 
\begin{equation}
\hat{b}_j=\sum_{i=1}^{n_{image}} a_{ij}x_i.
\end{equation}
 Similarly the backprojection accumulates in each element of the image $x_i$ the value from the detector considering the intersection length as
 \begin{equation}
 \hat{x}_i=\sum_{j=1}^{n_{detector}} a_{ij}b_i.
 \end{equation}
For most realistic x-ray tomography problems, the matrix $A$ is extremely large  and it is therefore  common not to pre-compute and store the matrix $\mathbf{A}$, but to compute the projection and the backprojection directly, by calculating the elements of $A$ on the fly when needed\cite{TIGRE}\cite{ASTRA}. This involves a large amount of simple, highly parallelisable arithmetic operations, which can be computed very efficiently using GPUs. Whilst this approach has been widely studied for voxels, tetrahedra require a carefully tuned algorithm to avoid numerical errors whilst remaining computationally efficient.

\subsection{Structuring the mesh}
\begin{figure}[htb]

  \centering
  \centerline{\includegraphics[width=5.5cm]{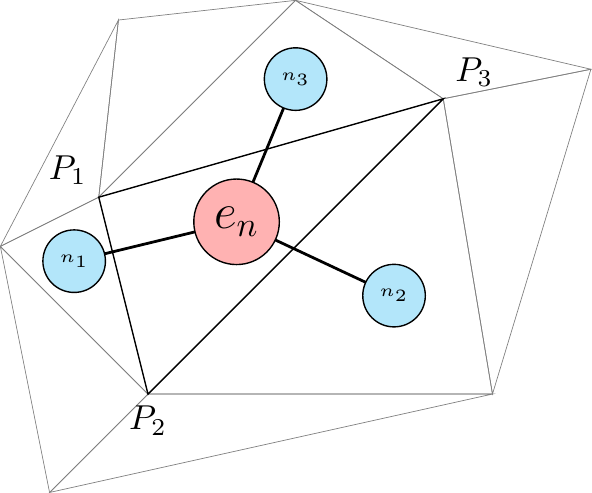}}
\caption{Graph representation of a triangular mesh in 2D. Each element $e_n$ contains three nodes $[P_1,P_2,P_3]$ and three neighbours $[n_1,n_2,n_3]$. The graph can be similarly built for 3D meshes.}
\label{fig:graph}
\end{figure}

Tetrahedral meshes describing a detailed geometry can be dense and the unstructured nature of common mesh representation is a challenge for tomography. In an unstructured mesh, every element has to be checked for intersection with each ray, which would increase the computational time by several orders of magnitude compared to the computation of  the non-zero intersections in a voxel-based methods, which can be computed very efficiently. While the common representation of meshes is unstructured, there is some structure also in general meshes and more advanced mesh representation are available that allow us to more easily exploit this structure. In this work, a graph-based representation of the mesh is used, as shown in figure \ref{fig:graph} (for the 2D version). The graph is constructed of individual elements that each contain $nD+1$ nodes $P$ and $nD+1$ ordered neighbour indexes $n$. This representation of the mesh, while slightly more memory consuming, provides neighbourhood information and thus allows for much more efficient algorithms for ray-tracing.

In addition to the graph containing both, element and neighbourhood information, our method uses a list of those elements that bound the triangulated space in order to allow us to efficiently find the first ray-tetrahedron intersection without the need for an exhaustive search (see section \ref{sec:init}).

\subsection{Tetrahedron-Ray Intersection}

\begin{figure}[htb]

  \centering
  \centerline{\includegraphics[width=3.5cm]{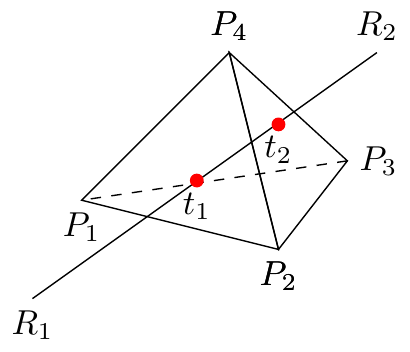}}
\caption{The intersection algorithm computes $t_1$ and $t_2$ and the face labels denoting the faces that are intersected.}
\label{fig:intersection}
\end{figure}

A fundamental step of the algorithm is the tetrahedron-ray intersection method that computes the length of the ray within the tetrahedron, which is essentially four triangle-ray intersections as seen in figure \ref{fig:intersection}. As a graph representation of the mesh is used, the index of the intersecting face of each tetrahedron can also be used to propagate the algorithm to the next intersecting element. This method however is not numerically robust when standard intersection methods are used and safeguards need to be added as discussed below.

Ray-triangle intersection algorithms have been widely studied, as the computer graphics field requires fast intersections for rendering with ray-tracing. The most common method is the M{\"o}ller Trumbore\cite{moller} algorithm as it is one of the fastest method available. A faster algorithm is available by Baldwin and Weber\cite{fastraytriangle}, however this requires precomputing and storing more data per triangle, which consumes considerably more memory in a volumetric triangulation. Both of these algorithms are based on computing the intersection location in barycentric coordinates and checking if the parameters lie inside the defined triangle. These methods however, have ``leaks'', as the techniques to reduce computation in the barycentric coordinate system can lead to misdirected or misrejected intersections, due to the discrete nature of the IEEE-754\cite{ieee754} floating point representation of decimal numbers. Watertight algorithms have been proposed in the literature to solve this problem\cite{watertight}. Watertight methods ensure that the numerical errors from the floating point arithmetic always lead to a detection as an intersection of an edge ray, which is the most desirable behaviour in computer graphics.

\begin{figure}[htb]

  \centering
  \centerline{\includegraphics[width=3.5cm]{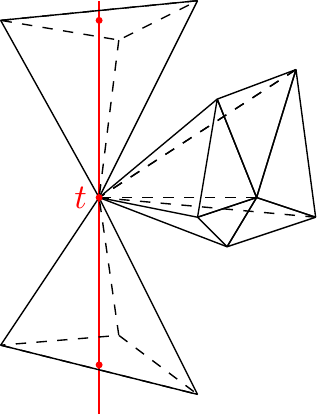}}
\caption{Tetrahedron structure that can cause issues in the ray propagation algorithm. Assume a full convex triangulation of the domain (we here only show a selection of tetrahedra for visualization purposes). As the elements at the top and bottom are not neighbours, the algorithm propagates through zero-length intersections. If the method is not watertight, there is a risk of not finding neighbours, whilst if the method detects too many intersections (too watertight), it risks looping though neighbours infinitely.}
\label{fig:bad}
\end{figure}

Both of these scenarios are undesirable in tomography. If an intersection is not detected due to leaks, the ray cannot propagate to the neighbouring element and raytracing stops prematurely. While detection of halted rays is simple, recovering from it can only be done successfully using a brute-force search of all existing tetrahedra. An example of a case where this would happen can be seen in figure \ref{fig:bad} where a ray goes trough a node of the element. It is possible that the intersection $t$ has not been detected due to precision errors and none of the neighbouring tetrahedra are guaranteed to detect the intersection either. The propagation algorithm would be stuck, and the only way of continuing would be to check intersections with all elements in order to find the closest intersection parameter $t$ to the latest valid one prior to the leak. This doesn't however ensure that tetrahedra will not be missed. 

The alternative watertight algorithm poses a different problem. Using figure  \ref{fig:bad} again as an example, a watertight algorithm would ensure that all tetrahedra touching the node are detected as intersecting with the ray. However, as all tetrahedra around the node are accepted intersections for propagation, the algorithm can get into an infinite loop. A list of intersected tetrahedra would need to be stored, but in parallel computing each processor would need to keep such a list and memory consumption would exceed realistic limitations.

The solution proposed in this article is using the M\"oller-Trumbore algorithm with an extra safety parameter that triggers if two intersections are not found in a tetrahedron. This safety parameter is applied in a way that effectively increases the size of the triangle faces. As the triangle increases size the intersection falls towards its interior, being safely detected as an intersection and keeping the value of the intersection parameter $t$ unchanged. The updated method can be seen in algorithm \ref{alg:moller}, where we introduce the parameter $\epsilon$. This method is not suitable to all ray-triangle applications, but works for our application. If an element is checked for intersection, it guarantees that it is already known to be intersecting once, therefore increasing the triangle marginally will not have the adverse effect of accidentally including false positives. It is however important to trigger the safety parameter only when less than two intersections are found and that the safety parameter $\epsilon$  is increased gradually, as otherwise the method would fundamentally become a less accurate version of the watertight algorithm.

\begin{algorithm}
\caption{M\"oller Trumbore with safety parameter $\epsilon$. $R_{\{1,2\}}$ define the ray and $P_{\{1,2,3\}}$ the triangle.}\label{alg:moller}
\begin{algorithmic}
\Require $R_1,R_2,P_1,P_2,P_3,\epsilon$
\State $\vec{d} \gets R_2-R_1 $
\State $\vec{E_1} \gets P_2-P_1 $
\State $\vec{E_2} \gets P_3-P_1 $
\State $\vec{q} \gets \vec{d}\times\vec{E_2}$
\State $a \gets \vec{E_1}\cdot \vec{q} $
\If {$a>-10^{-8}$ \textbf{and}  $a<10^{-8}$} \Comment Check if its zero
\State \Return \textbf{false}
\EndIf
\State $f \gets 1/a$
\State $\vec{s}\gets R_1-P_1$
\State $u \gets f (\vec{s}\cdot\vec{q})$
\If {$u < (0-\epsilon)$}
\State \Return \textbf{false}
\EndIf
\State $\vec{r} \gets \vec{s}\times \vec{E_1}$
\State $v \gets f(\vec{d}\times\vec{r})$
\If {$v < (0-\epsilon)$ \textbf{or} $(u+v)>(1+\epsilon)$}
\State \Return \textbf{false}
\EndIf
\State $t \gets f (\vec{E_2}\cdot\vec{r}) $
\State \Return  \{\textbf{true}, $t$\}
\end{algorithmic}
\end{algorithm}

\subsubsection{Floating Point Precision}\label{sec:fp}

GPUs do not handle double precision floating point arithmetic fast, even those designed specifically for double precision operations remain twice as slow. Memory also plays a role. Not only does double precision require twice the memory a more important bottleneck in GPU computing tends to be memory reading and waiting times. Therefore, most arithmetic on GPUs tends to use single precision operations. This is not true for the presented algorithm. Due to the scale differences involved in tomography, it is imporatnat that the intersection code has higher precision than the data that it uses. In IEEE-754 single floating point representation, the rounding error can be roughly estimated to be in the 8th or 9th digit after the most significant decimal point. However, the order of magnitude of $\vec{d}$ and $\vec{E}_{\{1,2\}}$ in algorithm \ref{alg:moller} can differ by the same amount. Even when the magnitudes are closer, the arithmetic operations can result in errors big enough that intersections can be mislabelled, specially when the triangles have high aspect ratios. An experiment showcasing this issue is presented in section \ref{sec:fp_results}.

\subsection{Numerically Robust Ray-Propagation}\label{sec:Algo}

The projection and backprojection algorithms are essentially the same, with the only difference that the projection operator integrates over the path into a detector pixel, and the backprojection operator gathers the detector pixel values into the element attenuation coefficients. Algorithm \ref{alg:main} describes the ray-tracing method. The algorithm essentially computes the current elements' intersection length, updates the integral (or the element, in  the case of the backprojection) and propagates to the element neighbouring the last intersection ($t_2$) plane. When the tetrahedon-ray intersection is checked there is a safety check to ensure that two faces are intersected and if not, then the safety parameter $\epsilon$ is increased, until two intersections are found. While this is a very rare event, the amount of tetrahedra and rays that tomography requires makes it statistically likely to happen. In some of our experiments this happened as rarely as five times every million rays, however due to the ill-posed nature of X-ray reconstruction, this event has noticeable negative effects on reconstruction. An extra check thus needs to be performed when zero-length intersections are found, to ensure there is no backtracking by choosing the wrong face for the propagation of the ray, which can happen when a node exist with several tetrahedra (see Fig.\ref{fig:bad}). 

Our algorithm has only one constraint: the volumetric mesh must be convex, however, as non-convex meshes can be convexified by the introduction of additional triangles, this is not very restrictive.

\begin{algorithm}
\caption{Robust ray-tracing for tetrahedon X-ray projection}\label{alg:main}
\begin{algorithmic}
\Require Geometry information, graph
\State \textbf{Launch Kernel} for every pixel $p[i,j]$ in the detector
\State $l\gets \lVert R2-R1 \rVert$
\State $\epsilon\gets 10^{-9}$, \hspace{10pt}$\sum \gets 0$ 
\State Read initial intersection $element$ index, $i_{now}$
\State \Return \textbf{if} $i_{now}=-1$

\While {$i_{now}\neq-1$}

\While {not Intersection}
\State $t_1,t_2 \gets$ TetraRayIntersection($i_{now}$, $\epsilon$)
\State $\epsilon \gets \epsilon\cdot 10$
\EndWhile
\State $\epsilon\gets 10^{-9}$
\State $\sum \gets \sum+l\cdot(t_2-t_1)\cdot element[i_{now}]$
\State if $t_2=t_1$ check if they need to be swapped
\State $i_{now}\gets$ neighbour of face where $t_2$  happened

\EndWhile
\State  $ p[i,j]\gets \sum$ 
\end{algorithmic}
\end{algorithm}

\subsection{Initialization of the Propagation Algorithm}\label{sec:init}

The previous section ignores an important issue: finding the first intersecting tetrahedron to initialzie the ray-propagating algorithm. To limit the memory usage and simplify the propagation code, Algorithm \ref{alg:main} requires the index of a tetrahedron on the mesh boundary.

The issue with initializing arbitrarily shaped convex tetrahedra meshes is that a brute force approach is not possible. Even for optimised meshes, with large boundary elements, the number of tetrahedra on the boundary of the mesh is typically larger than the number of times the X-ray path intersects with internal tetrahedra so that the initialization could take significantly longer than the X-ray propagation. On larger meshes, brute force initialization would thus dominate  computation time. The only viable solution is to avoid checking most of the boundary elements by implementing an efficient search strategy.

Quinto el. al.\cite{Quinto2013TetrahedralVR} propose a quadtree representation of the boundary surface triangles to initialise  rays propagation. This approach works very efficiently, but adds a significant topological constraint to the input mesh: it requires that surface elements are arranged on a regular mesh, with edges aligned to the quadtree axes. This implies that the shape and size of the elements are strongly constrained. As the proposed method in this article aims for a generic implementation with minimal constraints on the input mesh shape, an alternative modified approach is proposed. 

In this work, an R*-tree\cite{beckmann1990r} is precomputed for the boundary elements in a pre-processing step in order to accelerate the initialization procedure. An R*-tree is a depth-balanced search three that is constructed by bounding regions containing each nodes children. They are a variation of R-trees, search trees designed to contain volumetric objects, optimised for spacial access, i.e. to pack objects that are closer together by some metric. The R*-tree is a variation to the standard R-tree\cite{guttman1984r} which minimises not only the area of each node, but also the overlap between nodes. These search trees are particularly interesting for the initialization of X-rays as the depth balance of R*-trees guarantees similar computational costs for all X-ray paths, which is beneficial in parallel implementations. Additionally, the box-shaped bounding regions that define each node allow for very fast box-ray intersection checks. 

To search the tree, a depth-first algorithm is implemented, as it requires minimal memory allocation per search, which can be a critical factor for GPU implementation. When a leaf node is reached in the search, all tetrahedra within that node are checked for intersection and the index of whichever has the minimum intersection parameter, i.e. the earliest intersection along the X-ray path, is stored.

R*-trees have a minimum and maximum number of children in each node, which implicitly defines the depth for a given input database. We have chosen 10 as maximum number elements and 4 as minimum, as it generates trees that are not too deep, but containing a small number of tetrahedra in each leaf node, which minimizes the number of tetrahedra that need to be checked for intersection by the depth-first method.

\subsection{GPU Implementation Details}

In GPUs, paralellization happens in small blocks of execution threads. Each thread inside a blocks is executed in parallel (not necessarily concurrently) and the GPU waits until the entire block is done before allocating a new one to the processor cores responsible for the execution. Thus, ensuring that all threads within a block execute similar code and terminate at the same time increases overall computation speed, as it minimises idle threads.

In tomography, the highest cost comes from memory reads, as they can be two orders of magnitude slower than an arithmetic operation and lead to idle threads waiting for read operations to be completed. Therefore, most of the optimizations that happen in GPU code relates to executing code that would access the same memory at the same time, to increase cache hits. Tetrahedron based tomography however is not very sensitive to this optimization, as the unstructured nature of the mesh means that even adjacent rays are not necessarily reading the same location in memory, i.e. they are not crossing the same tetrahedra nor the same amount of tetrahedra. On top of that, the nature of the algorithms proposed here means that they require several branches per execution block, so thread divergence is high (an undesired behaviour on GPU parallelization). It is  likely that thread divergence hides memory latency problems as the fastest speed can be achieved by computing the rays in small square blocks. We found empirically that $8\times8$ blocks works best on GTX 10XX GPU models. We suggest empirically testing different size if this code is required to execute on alternative GPUs.

The backprojection is more problematic. When launching the algorithm in parallel, multiple threads want to update the same attenuation coefficients. Atomic operations ensure that no race conditions are met, but in theory this increases the time significantly as most of the threads could be waiting to write. While separating the rays may seem the best approach, empirical tests show that running the backprojection similarly to the projection gives the best results. This unintuitive behaviour is likely caused by the thread divergence as it can hide atomic writing latency. On top of that, the general nature of the algorithm and tetrahedra meshes can result in a limited amount of simultaneous write attempts, depending on the particular input mesh.

For our multi GPU approach, projections are divided, while keeping the full mesh in each of the GPUs memories. This is because dividing the graph in pieces of similar computational cost that are also convex is a significant challenge, and its size would generally be small enough to fit entirely in compute oriented GPUs.

\section{Results}
\label{sec:results}

To evaluate the quality and performance of the algorithm initial experiments are presented here.

\subsection{Image Reconstruction}
\begin{figure}[htb]
  \centering
  \centerline{\includegraphics[width=6cm]{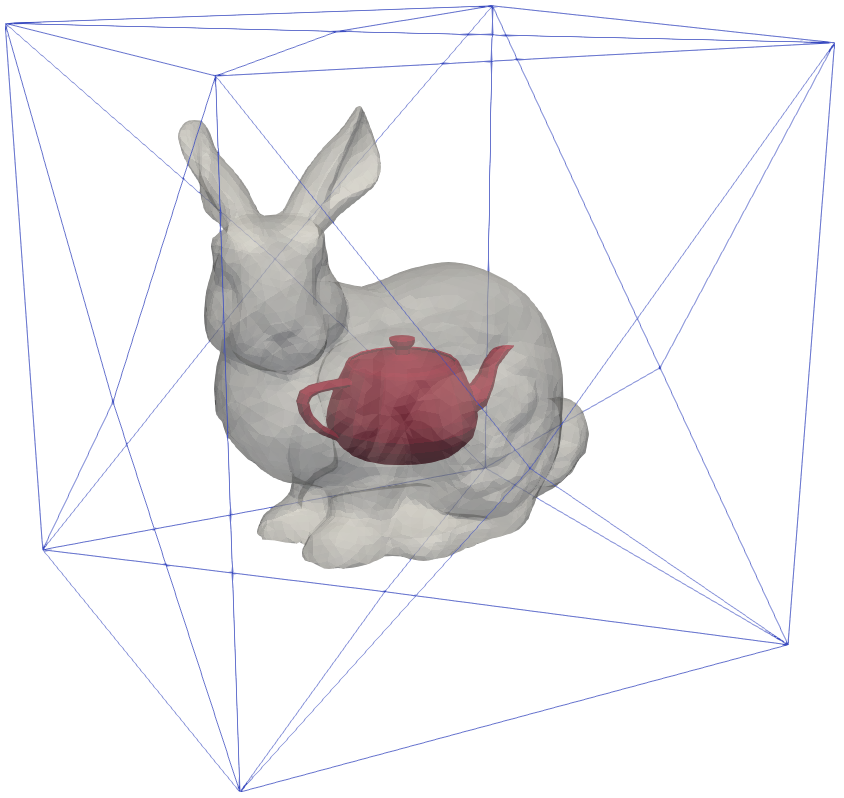}}
\caption{Surface representation of the tetrahedra mesh used for validation of the forward and backward projector.}
\label{fig:bunnydata}
\end{figure}

To test the tetrahedral mesh projector and backprojector a simulated experiment is presented. Using a tetrahedral mesh, X-ray projections are simulated and reconstructed. We here use our new tetrahedral approach and contrast it to standard voxel based reconstruction using the TIGRE toolbox\cite{TIGRE}. The surface representation of the data used is shown in figure \ref{fig:bunnydata}. The mesh is generated using three surfaces: a cube bounding the volume area, a closed surface version of the Stanford bunny and the Utah teapot. The models are meshed using the Simpleware ScanIP, which generates a wide rate of tetrahedral sizes and provides individual tetrahedra with high aspect ratios. The resulting mesh contains approximately 175,000 tetrahedra. Attenuation coefficients of values 0, 1 and 2 are assigned to the tetrahedra in the box, the bunny and the teapot respectively.

\begin{figure}[htb]

\begin{minipage}[b]{.48\linewidth}
  \centering
  \centerline{\includegraphics[width=4.0cm]{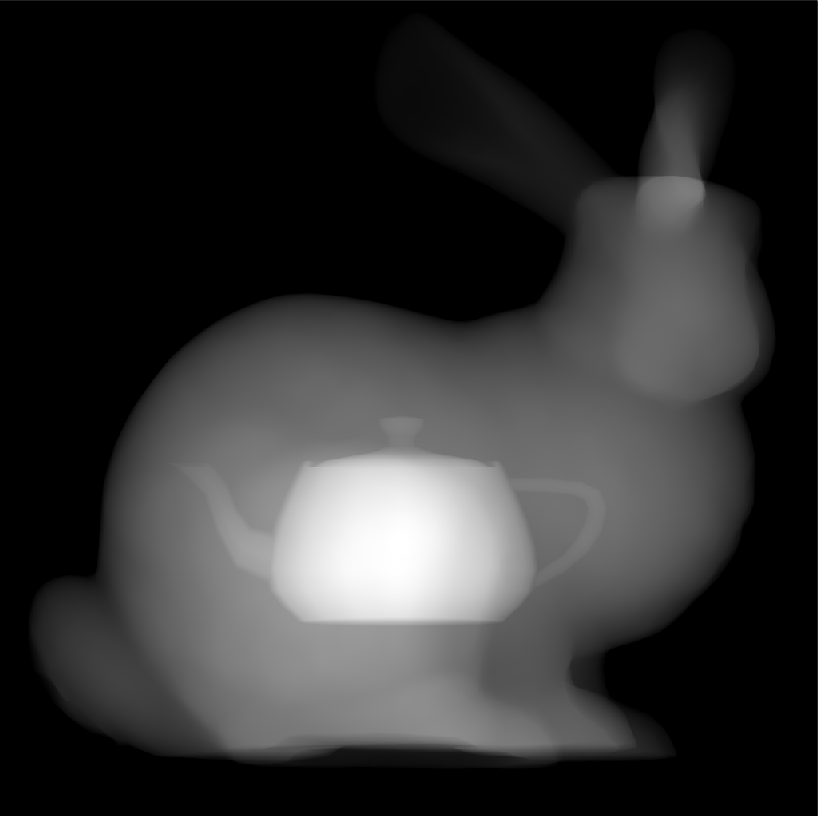}}
\end{minipage}
\hfill
\begin{minipage}[b]{0.48\linewidth}
  \centering
  \centerline{\includegraphics[width=4.0cm]{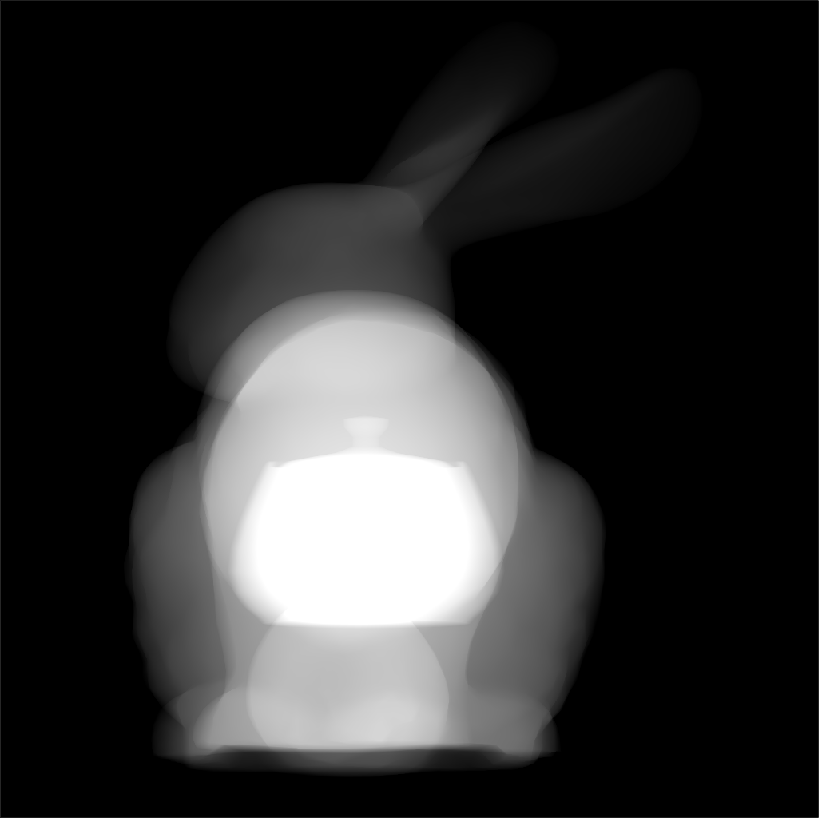}}
\end{minipage}
\caption{Simulated X-ray projections for $0^{\circ}$ and $90^{\circ}$ rotation positions.}
\label{fig:proj}

\end{figure}

Using the  tetrahedra based forward projector, 100  projections of \mbox{$1024\times1024$} pixels are simulated around a circular trajectory in equidistant angles for a cone beam. Figure \ref{fig:proj} shows projection at angle $0^{\circ}$ and $90^{\circ}$.

\begin{figure}[!htb]

\begin{minipage}[b]{.48\linewidth}
  \centering
  \centerline{\includegraphics[width=4.0cm]{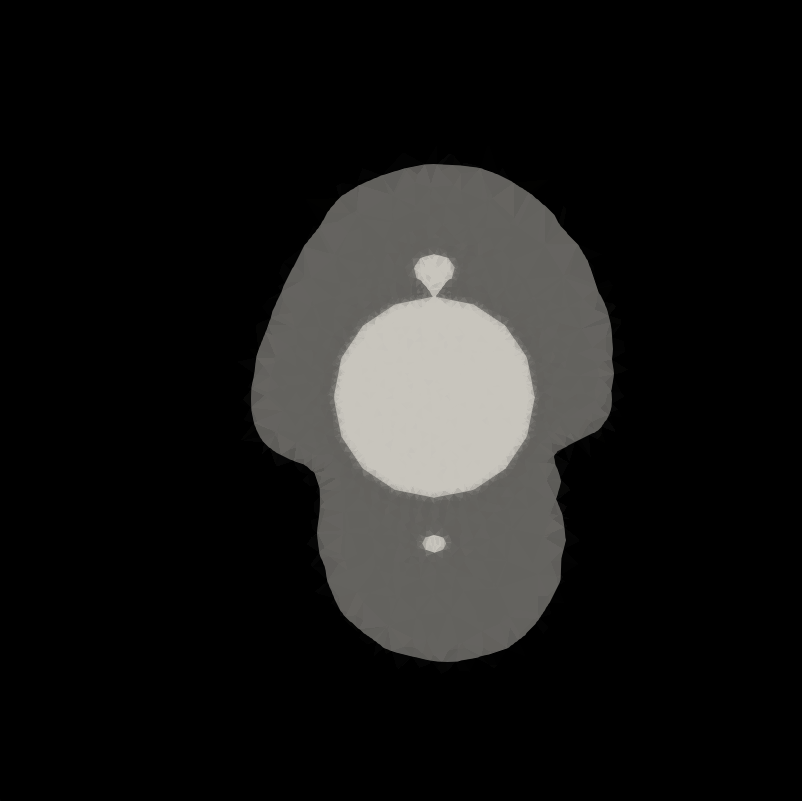}}
\end{minipage}
\hfill
\begin{minipage}[b]{0.48\linewidth}
  \centering
  \centerline{\includegraphics[width=4.0cm,trim={1cm 1cm 1cm 1cm},clip]{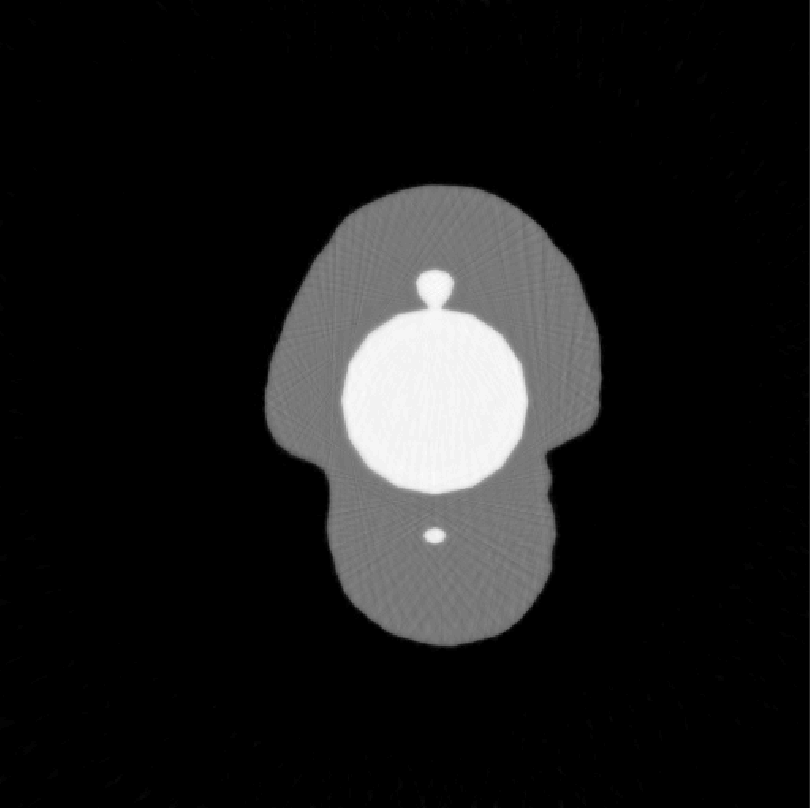}}
\end{minipage}

\vspace{0.1cm}

\begin{minipage}[b]{.48\linewidth}
  \centering
  \centerline{\includegraphics[width=4.0cm]{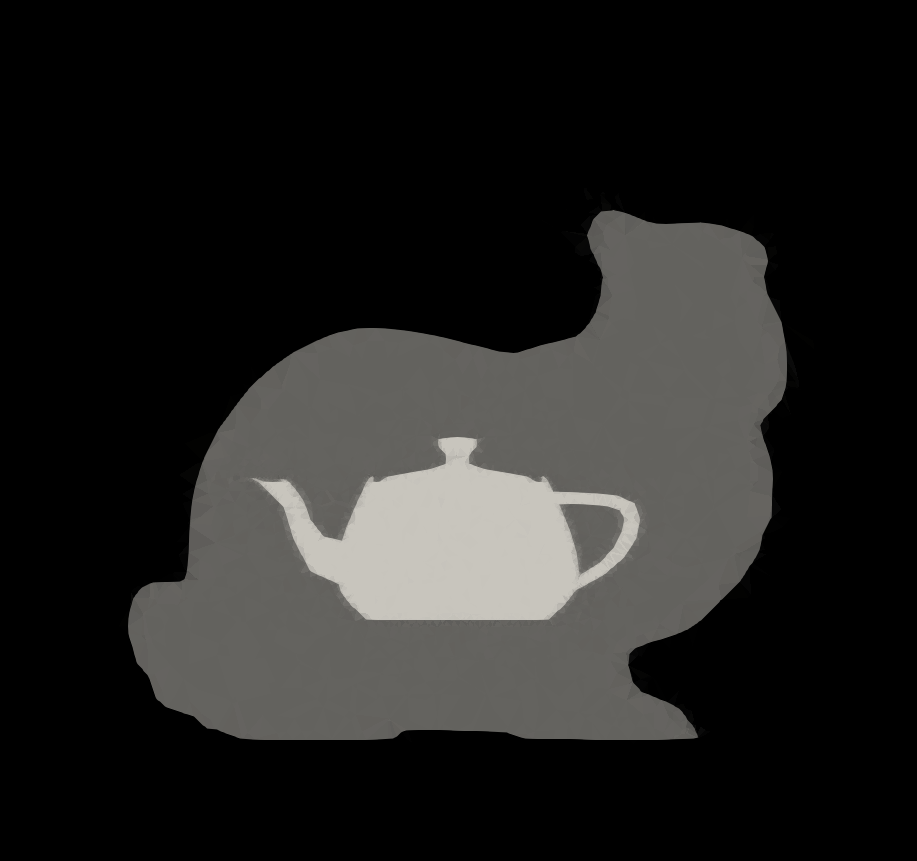}}
  \vspace{0.3cm}
   \centerline{(a) Tetrahedra basis}
   \centerline{on known mesh}\medskip
\end{minipage}
\hfill
\begin{minipage}[b]{0.48\linewidth}
  \centering
  \centerline{\includegraphics[width=4.0cm,trim={1cm 2cm 1cm 1cm},clip]{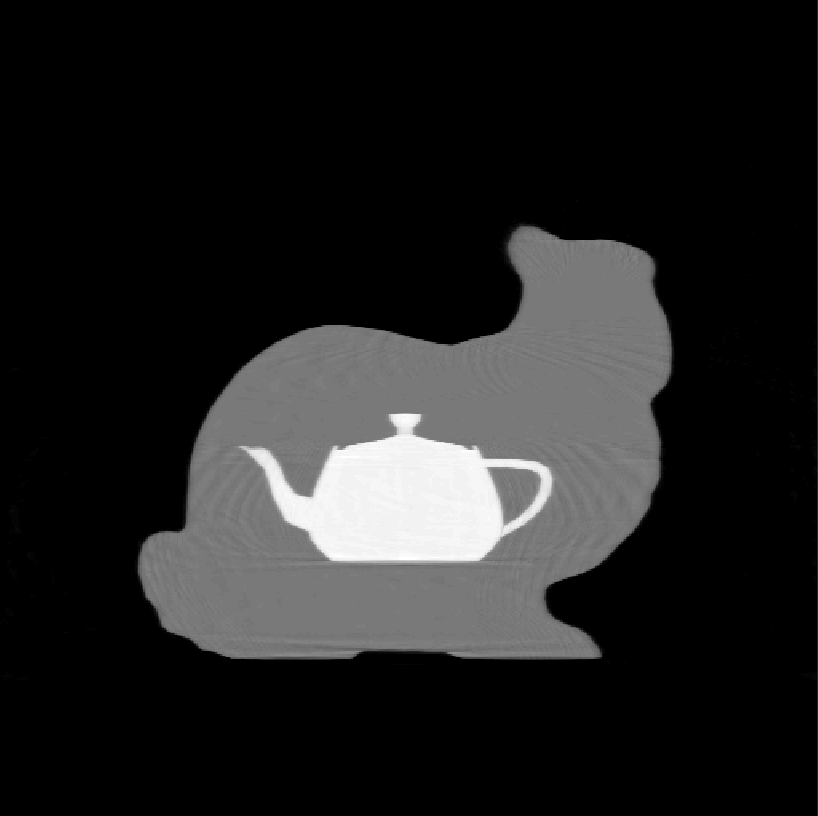}}
  \vspace{0.3cm}
    \centerline{(b) Voxel basis}
\centerline{in high resolution}\medskip
\end{minipage}
\begin{minipage}[b]{.48\linewidth}
  \centering
  \centerline{\includegraphics[width=4.0cm]{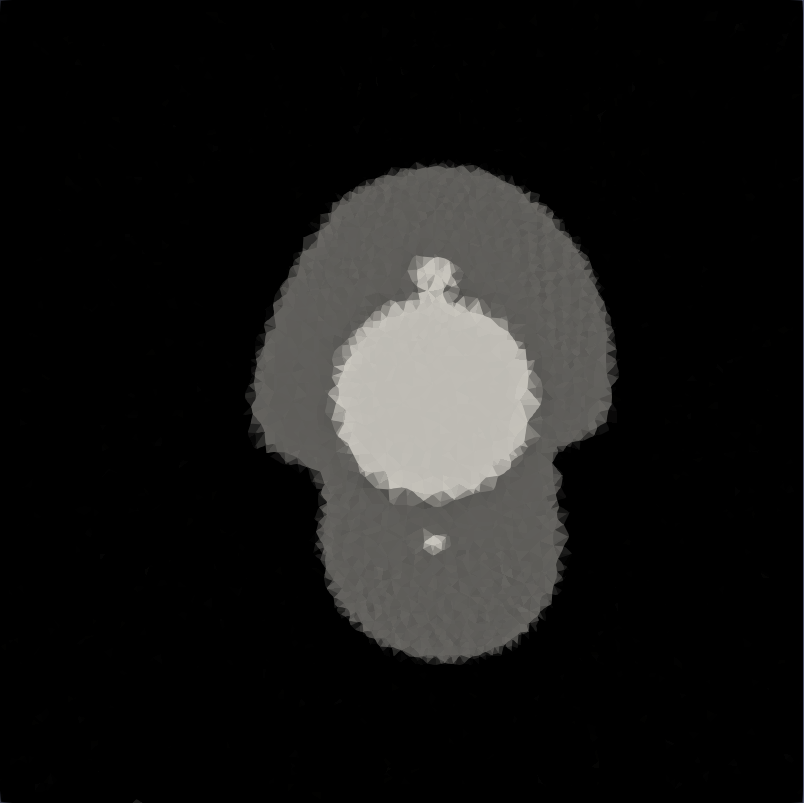}}
\end{minipage}
\hfill
\begin{minipage}[b]{0.48\linewidth}
  \centering
  \centerline{\includegraphics[width=4.0cm,trim={0.6cm 0.6cm 0.6cm 0.6cm},clip]{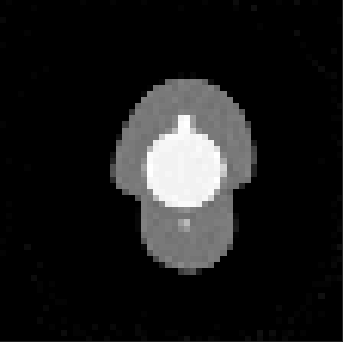}}
\end{minipage}

\vspace{0.1cm}

\begin{minipage}[b]{.48\linewidth}
  \centering
  \centerline{\includegraphics[width=4.0cm]{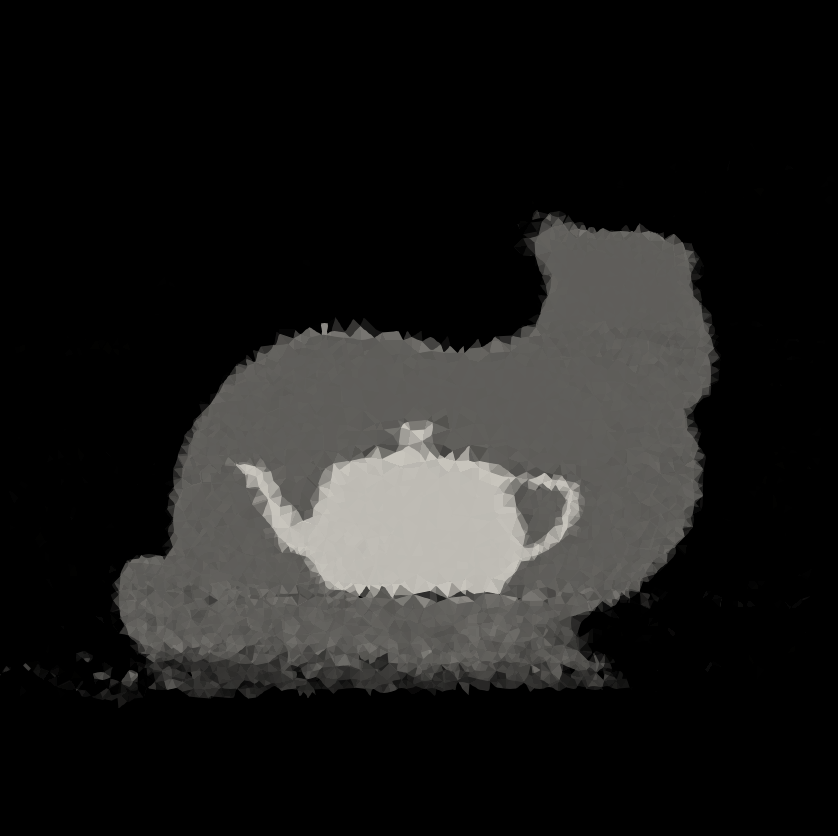}}
  \vspace{0.3cm}
   \centerline{(c) Tetrahedra basis}
   \centerline{on unknown mesh}\medskip
\end{minipage}
\hfill
\begin{minipage}[b]{0.48\linewidth}
  \centering
  \centerline{\includegraphics[width=4.0cm,trim={0.6cm 0.6cm 0.6cm 0.6cm},clip]{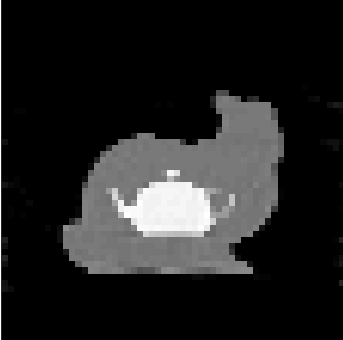}}
  \vspace{0.3cm}
    \centerline{(d) Voxel basis}
\centerline{in low resolution}\medskip
\end{minipage}

\caption{Iterative reconstruction using the OS-SART algorithm with 50 iterations, block size of 20 projections with 100 projections in total. Reconstruction is shown in (a) tetrahedra basis with a known mesh (170k tetrahedra) and (b) voxels ($512^3$ voxels), (c) tetrahedra basis on unknown mesh (3500k tetrahedra) and (d) voxel basis on low resolution ($56^3$ voxels, the same amount as tetrahedra are in (a)). The attenuation coefficients are shown in range \mbox{[0-2.1]}}
\label{fig:rec}

\end{figure}

Four reconstructions are computed. On a tetrahedra basis reconstructions, the same mesh as the one used to generate the data is used, as well as a ``high resolution'' 3500k element mesh. For the voxel based representation a $512\times 512\times 512$ mesh, and a $56\times 56\times 56$ mesh (the same amount of elements as the original tetrahedra mesh) spanning the same volumetric area is used.  OS-SART\cite{OS_SART} is here used for reconstruction. The algorithm is run with the same parameters for both reconstruction volumes, using 50 iterations with blocks of 20 projections. Cross section results of the reconstruction are shown in figure \ref{fig:rec}.

Comparing reconstructions is non-trivial, as any comparison would necessarily need to map from one of the basis functions used to the other one, likely obfuscating benefits or errors that each of the basis has. For example, converting the voxel basis to tetrahedra basis for comparison may hide the blur on boundaries or the visible streak artefacts on the reconstruction. On the other hand, converting from the tetrahedra basis to voxel may show that the mesh-based model reconstructs images that are more uniform, but this is just a consequence of a particularly large tetrahedra on a given experiment, not necessarily due to general robustness. However, one can observe that the nature of the errors are significantly different between the two image types. The voxel basis image shows typical streak artefacts for low angle scans or horizontal flat surfaces and a general blur, especially around small features. The tetrahedral basis image however does not seem to be affected by the streak artefacts when the surface mesh is known. It does however, accumulate most of the errors around the boundaries of objects.

Note that we here find the best reconstruction using tetrahedra basis using the same mesh to generate the projection data and compute the tetrahedral reconstruction. Whilst this is not realistic in real applications, it is done here to 1) show that our method does accurately compute forward and backward projections and thus provides low error reconstructions in the ideal case and 2) to demonstrate that tetrahedral meshes can have advantages over voxel basis if a good mesh is chosen, as the voxel basis has more unknown elements than observations leading to typical artefacts, whilst the mesh representation has fewer unknowns than measurements and thus does not suffer from the same problem if a good mesh is available. Obviously finding a good mesh is in itself an important issue, as can be seen in Figure 6 (c), where a random mesh seems to enhance the streak artefacts visible in the voxel meshes. 


\subsection{Image Reconstruction of a CAD model}
\label{sec:unknownmesh}
\begin{figure}[!htb]

  \centering
  {\includegraphics[width=6cm]{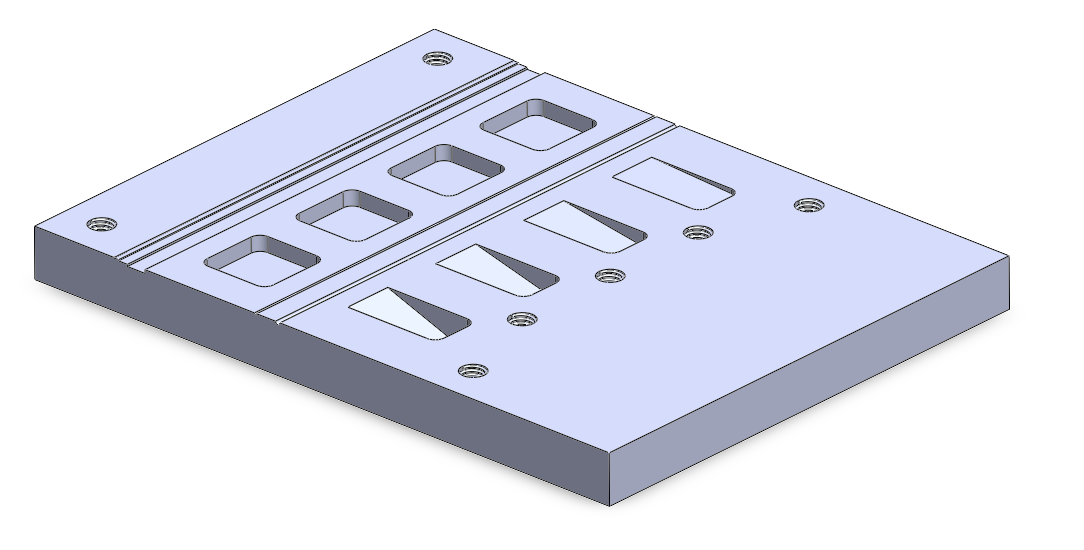}}
  {\includegraphics[width=6cm]{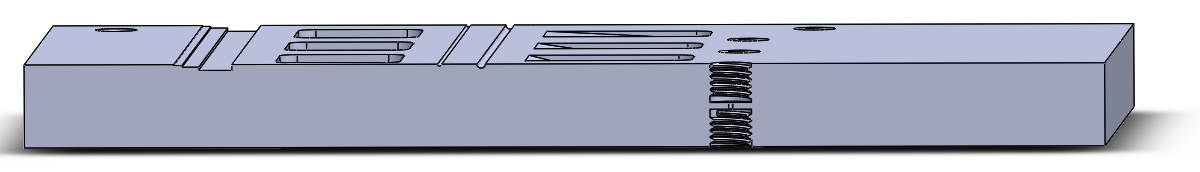}}

\caption{CAD model designed for CT image quality and feature detection evaluation.}
\label{fig:CAD}

\end{figure}

In tetrahedral mesh tomography, the structure and shape of the tetrahedra can have mayor importance on the quality of the reconstruction. Approaches to refine the a mesh have been published (e.g. by Yamanaka et al.\cite{Yamanaka2013} for tetrahedra reconstruction with FDK) whilst solutions to align available CAD models to projection measured data also exist\cite{Ito}. The following experiment highlights the difference between having accurate knowledge of a prior model and having an ideally aligned mesh generated by a prior model. We use the CAD model in Figure \ref{fig:CAD} to simulate 100 projections of size $1024 \times 1024$. The same model has been used to generate a tetrahedra mesh with approximately 620k elements, while a tetrahedra mesh with 980k elements is also created of the same dimensions, but without knowledge of the object boundaries.

\begin{figure}[htb]

\begin{minipage}[b]{.48\linewidth}
  \centering
  \centerline{\includegraphics[width=4.0cm]{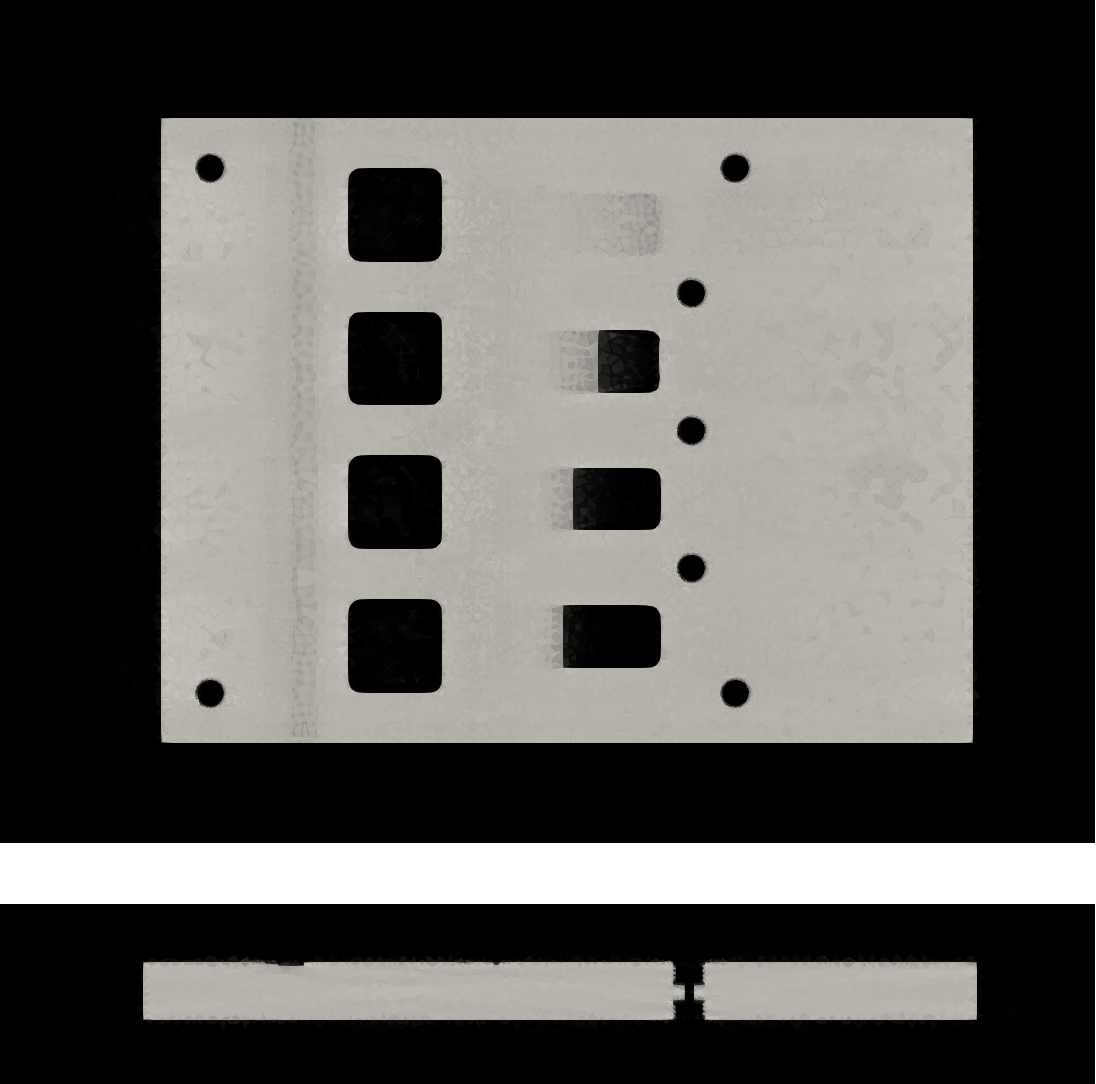}}
  \vspace{0.3cm}
   \centerline{(a) Mesh with prior information}\medskip

\end{minipage}
\hfill
\begin{minipage}[b]{0.48\linewidth}
  \centering
  \centerline{\includegraphics[width=4.0cm,clip]{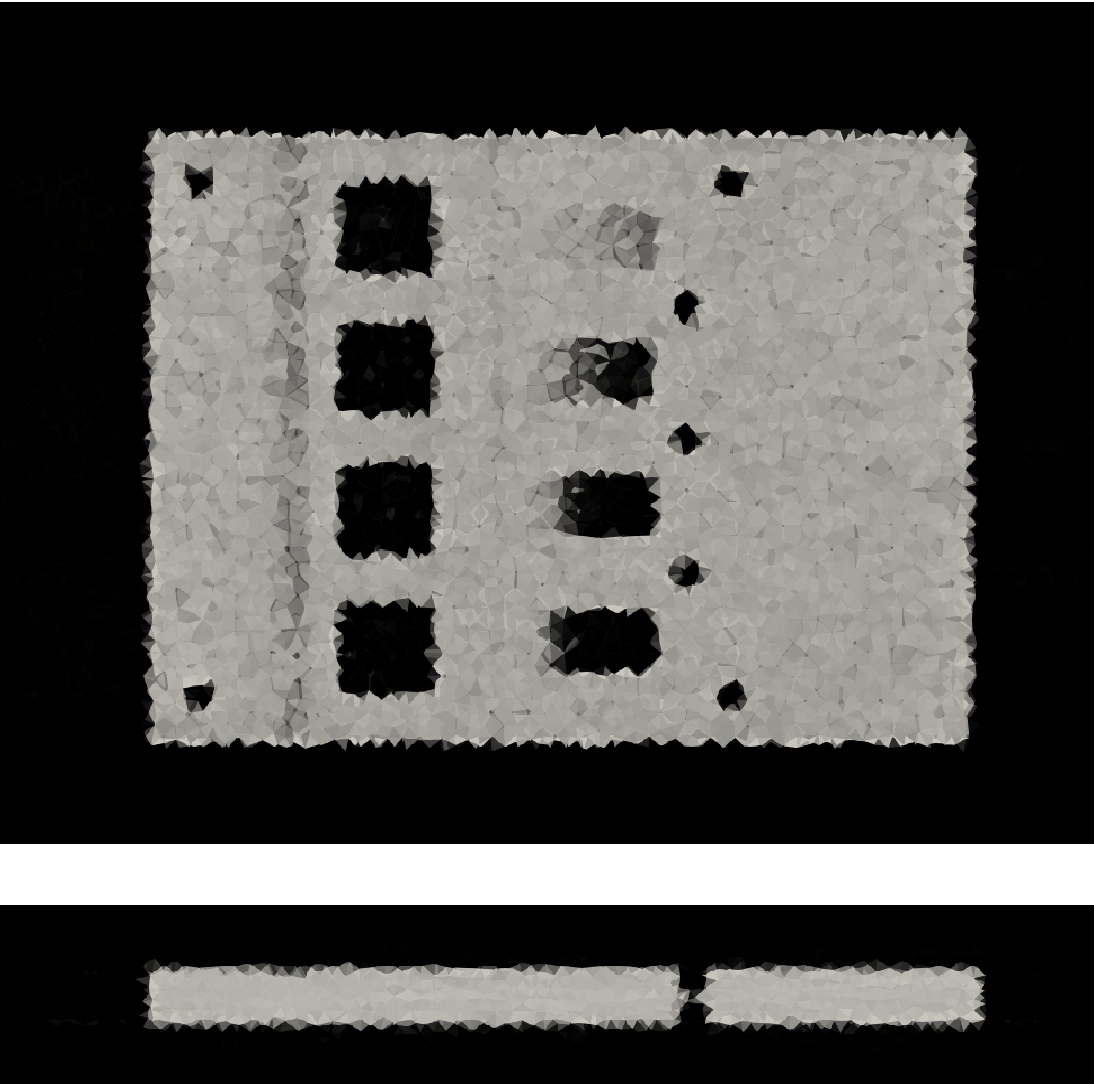}}
  \vspace{0.3cm}
    \centerline{(b) Dense random mesh}\medskip

\end{minipage}

\caption{Iterative reconstruction using the OS-SART algorithm with 50 iterations, block size of 20 projections with 100 projections  for tetrahedra mesh based images, (a) with prior surface information and (b) without prior information. The attenuation coefficients are shown in range \mbox{[0-1.1]}}
\label{fig:rec2}

\end{figure}

Figure \ref{fig:rec2} shows the reconstructionusing the two different meshes. It can be clearly seen that having prior information on the tetrahedral mesh can have major effects on reconstruction quality, even with meshes that are significantly smaller in terms of available elements.

\subsection{Floating point errors}\label{sec:fp_results}

\begin{figure}[htb]

\begin{minipage}[b]{.48\linewidth}
  \centering
  \centerline{\includegraphics[width=4.0cm]{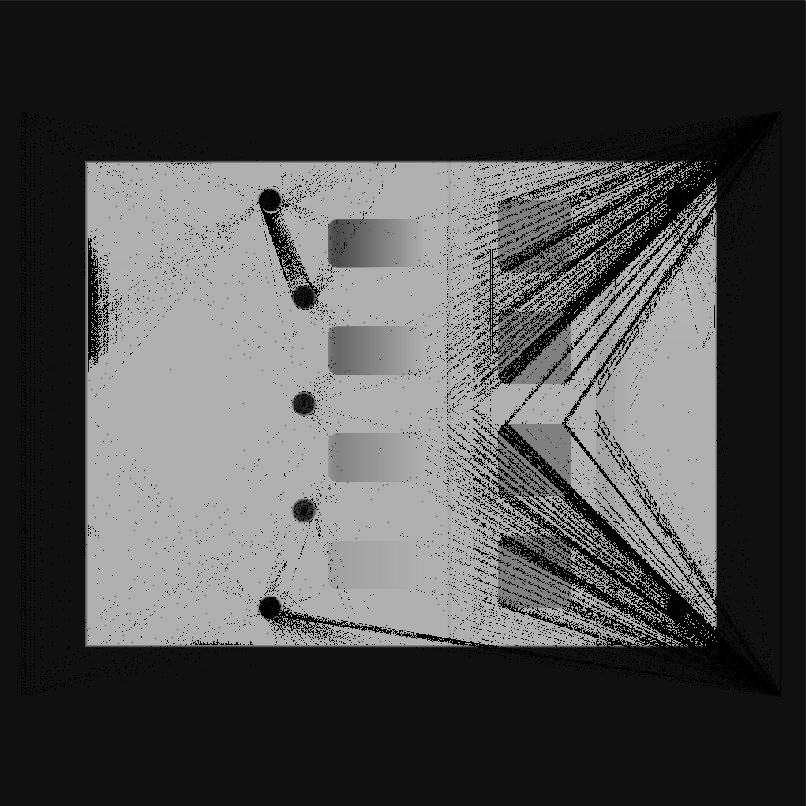}}
  \centerline{(a) Single precision}\medskip
\end{minipage}
\hfill
\begin{minipage}[b]{0.48\linewidth}
  \centering
  \centerline{\includegraphics[width=4.0cm]{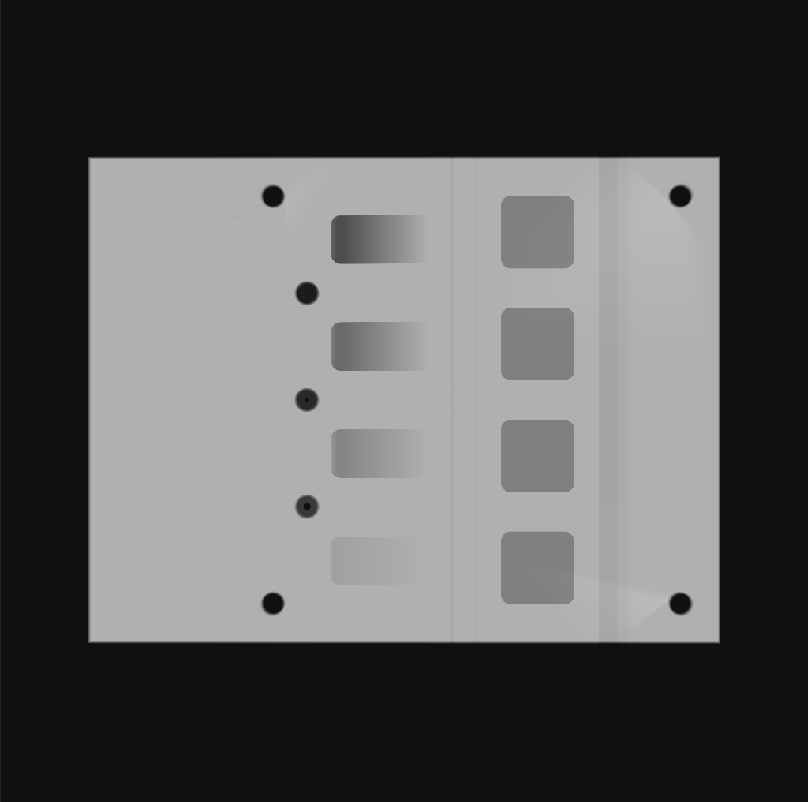}}
  \centerline{(b) Double precision}\medskip
\end{minipage}
\caption{Effect of floating point types in the ray-triangle intersection algorithm for a specific projection. Black dots in (a) are rays that force the algorithm to terminate abruptly or loop infinitely.}
\label{fig:singledouble}
\end{figure}

As  mentioned in section \ref{sec:fp}, using double precision floating point arithmetic (or, in general, higher precision numerics than the precision of the mesh) for the triangle intersection code is critical, especially in cases where the aspect ratio of the triangles is high or the triangles are very small. Figure \ref{fig:singledouble} shows a projection of a single material piece that has been triangulated with a Delaunay triangulation algorithm using the points of an STL file with added nodes on the edges to ensure surface preservation. This generates a high definition triangulation with the minimal number of tetrahedra, but with a low quality mesh \footnote{In terms of FEM mesh quality metrics} due to high aspect ratios. This type of model can be useful to simulate projections from known data e.g. from CAD models. Results in figure \ref{fig:singledouble}(a) demonstrates how using the single precision numerical intersection code results in a high number of pixels where the ray-propagation integrals fail to terminate properly, either by sudden termination of the propagation (due to missed intersections) or due to infinite looping through triangle neighbourhoods. 

\subsection{Performance figures}

Evaluating performance is not straightforward. The forward and backprojection algorithms are effectively the same, but the sub-algorithm used for initialization is fundamentally different from the ray propagation algorithm with both performing significantly difference. As arbitrary meshes can vary significantly in the number of boundary elements compared to the number of total elements, the overal algorithm performance can either be dominated by the initialisation (if there are relatively many boundary elements) or by the ray tracing (if there are relatively few boundary elements). Even if two meshes have the same number of boundary and internal elements, the shape of these elements can cause major differences in performance of each of the two sub-algorithms, making general performance figures  hard to derive. 

To provide some intuition into the performance, a particular type of regular mesh is thus here used as a benchmark. We generate voxel-type regular meshes that are then triangulated. This type of mesh does not exploit the capabilities of tetrahedra mesh reconstruction but allows for reasonable performance computation and compassion. Whilst these results are not directly usable to predict  performance of arbitrary meshes, an approximate estimation can be obtained by adding the results for initialization and ray-propagation for particular meshes after acounting for the n umber of boundary elements and total elements. For example, the mesh on Figure \ref{fig:rec2}(a) has $6\times 10^5$ total elements, but contained only 192 boundary elements.

\begin{figure}[htb]

  \centering
  \centerline{\includegraphics[width=8cm]{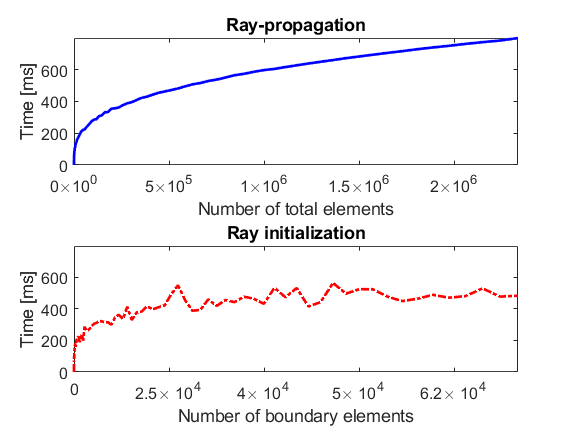}}
\caption{Kernel initialization and ray-propagation times for a single $1024 \times 1024$ projection at different mesh sizes and boundary sizes.}
\label{fig:time}

\end{figure}

Figure \ref{fig:time} shows computing times on a single GTX 1050 GPU for regular tetrahedra meshes of increased size. Both measurements should be interpreted separately. The regular meshes have been created by linearly increasing the number of points along each edge. The projection kernel linearly increases its computational cost with respect to  edge length. This behaviour is expected, as the ray-propagation kernel is expected to encounter a number of elements that linearly increases with edge length on a regular grid, but the measurement highlights that the algorithm does not have a memory or compute limitation related to mesh size. The initialization kernel also behaves as expected. It showcases more jagged lines, caused by differences in the R*-tree generation for that particular shape and size. We hypothesize that a better R*-tree bulk loading\cite{kamel1993packing} procedure may be able to generate a more stable profile, but have not tested this. The initialization procedure increases logarithmically with the edge length, which is also what is expected with tree-like search algorithms. 

Memory transfer speeds are not show as they almost always had the same speed, reaching 20 ms of total time for small sizes while keeping that same total time on the largest meshes. 

As the multi-GPU splitting method used is simple copying the entire memory to all GPUs and splitting the total projections between the available GPUs, no performance measures are required. The time per projection is the same, but now some projections are computed simultaneously. The minor overhead of multiple memory copies can be practically ignored due to its short length compared to total computation times. 

It is worth noting that voxel-based tomographic reconstruction is able to compute a single projection from approximately $2\times 10^6$ voxels (i.e. a $128^3$ voxel image) onto the same $1024\times 1024$ detecrtor in around 120 ms including memory transfer. The voxel-basis projection thus remains an order of magnitude faster than a tetrahedral mesh based projection. This speed advantage is due to the  regular structure in voxel based meshes. Nevertheless, using tetrahedral basis meshes enables the use of significantly smaller meshes to accurately represent images to the same level of detail and thus provides a valuable additional tool for tomographic reconstruction.

\section{Discussion}
\label{sec:discussion}

We developed a numerically robust, GPUs accelerated method for X-ray forward and backprojection on convex meshes with arbitrarily shaped tetrahedral elements. Numerical results using challenging meshes shows the robustness of the new approach.  We presented some comparisons to standard voxel based reconstructions, showing that better results can be obtained if good boundary mesh estimates are available.  Whilst computation speed for generic meshes remains larger than that for voxel based meshes with the same number of elements, tetrahedral meshes have significantly increased flexibility so allow meshes with significantly fewer elements to be used without a decrease in image quality. 

The focus here has been on the description of the algorithm and the demonstration of its computational capabilities. Having developed this new tool, there are now a range of interesting open questions that we are addressing in ongoing research. 

Our tool allows efficient simulation of transmission tomography imaging applications from mesh based representations. The method can for example be extended to generate accurate X-ray projections from multi-material volumes, for example using Monte Carlo photon simulations. Whilst most existing software is based on surface models, using our volumetric models allows for the specification of more gradually varying material changes with arbitrary surfaces.

Tetrahedra based image reconstruction also has a strong potential in image reconstruction applications where the sample geometry is known, as we already highlighted in \cite{initTetra}, where the mesh model seem to outperform voxel-based methods on limited angle scans thanks to the use of prior surface information. The next steps to test the mesh-based method is to use it together with algorithms that align prior CAD models to projection data and algorithms that refine tetrahedra meshes for surface characterization. Together with iterative reconstruction capabilities, these two methods have the potential to allow reconstruction from limited numbers of projections and fault detection be performed fast and reliably in in-process CT applications. Similarly, the tetrahedral reconstruction could be of benefit in image-based modelling applications, where a sample is scanned tin order to create a physical model to be simulated. These models are often based on finite element methods and thus require  a tetrahedral basis, which are currently generated using segmentation techniques on voxel-based reconstructions. By directly reconstructing with a tetrahedral basis, more accurate surfaces estimates may be possible. 
The code in MATLAB and CUDA can be freely accessed at \url{github.com/AnderBiguri/TriangleCT}.

\section*{Acknowledgement}
We gratefully acknowledge the support of NVIDIA Corporation with the donation of the Titan
Xp GPU used for this research. This research was supported by EPSRC grant EP/R002495/1 and
the European Metrology Research Programme through grant 17IND08. A. Biguri would like to thank Marco Vallario and Tobias Bertel for interesting and productive discussions on ray tracing and search trees.

\bibliographystyle{IEEEbib}
\bibliography{strings,refs}

\end{document}